\documentclass[showpacs,preprint,aps]{revtex4}
\usepackage{graphicx}
\usepackage{dcolumn}
\usepackage{bm}
\begin{document}
\setcounter{page}{1}
\title 
{Three body dwell time}
\author
{N. G. Kelkar}
\affiliation{ 
Departamento de Fisica, Universidad de los Andes, 
Cra.1E No.18A-10, Bogota, Colombia}
\begin{abstract}
The lifetime of an unstable state or resonance formed as an intermediate 
state in two body scattering is known to be related to the dwell time or the  
time spent within a given region of space by the two interacting particles. 
This concept is extended to the case of three body systems and a 
relation connecting the three body dwell time with the two body
dwell times of the substructures of the three body system 
is derived for the case of separable wave functions. The 
Kapur Peierls formalism is revisited to discover one of the first definitions
of dwell time in literature. An extension of the Kapur Peierls formalism to the
three body case shows that the lifetime of a three body resonance can indeed 
be given by the three body dwell time. 
\end{abstract}
\maketitle 
\section{Introduction}
Tunneling is one of the most exotic phenomenon in quantum mechanics 
and the study of tunneling times with its different controversial 
definitions is equally so. In an attempt to find out how long a 
particle needs to traverse a potential barrier, physicists gave rise 
to several definitions such as the dwell time, phase time, arrival time, 
Larmor time, traversal time, residence time and more abstract complex times 
\cite{times}. Of these, the dwell and phase times seem to be the most 
relevant for the study of unstable or resonant states occurring in 
tunneling as well as scattering problems.  The dwell time 
(sometimes known as the sojourn or residence time) is defined as the average 
time spent by a particle in a given region of space. The introduction of 
the dwell time concept is commonly attributed to F. Smith \cite{smith} in 
literature. Smith introduced it in connection with quantum collisions and 
B\"uttiker \cite{buet} discussed it in the context of one dimensional 
tunneling. However, it is interesting to note that long before in 1938 
\cite{kppl}, 
P. L. Kapur and R. Peierls had derived the formula for 
dwell time as we know it now. It was a by-product of the formalism for 
cross sections with resonances in nuclear reactions and they did not 
explicitly mention it to be a quantum time. In the case of tunneling 
through a barrier, the average dwell time is the time spent by the 
particle in the barrier, irrespective of the fact if it got transmitted 
or reflected. However, there do exist definitions of reflection and 
transmission dwell times which are connected to the measured lifetimes of 
decaying objects \cite{wealphagoto}. A review of some 
formal aspects of the dwell time distributions including the operator 
representation of the dwell time and its extension to free multiparticle 
systems can be found in \cite{adolfo}. 

The dwell time concept finds applications in various branches of 
physics. In \cite{sokol} for example, the author relates this concept to 
the qubit residence time measurements in the presence of Bose Einstein 
condensates. In another recent work \cite{dresel}, it is shown that as a 
result of the Dresselhaus spin-orbit effect, the difference between the 
dwell times of spin up and spin down electrons can become greater as the 
semiconductor length increases. These studies could be useful in 
designing quantum spintronic devices. Some other applications 
include studies with semiclassical theories of quantum chaos \cite{chaos} 
and the connection of dwell time with a quantum clock. 
Salecker and Wigner \cite{saleck} proposed a microscopic clock 
to measure distances between space time events. Peres \cite{peresclock} 
extended the formalism 
to the measurement of an average time spent by particles in a given region 
of space. C. R. Leavens \cite{leaven} established the connection between 
Peres's time spent in a given region of space and the standard definition of 
average dwell time. More recent discussions of the Peres clock and 
dwell times can be found in \cite{lunarpark}. 
Finally, the dwell time 
is useful in characterizing resonances \cite{portme, meprl} as 
well as studying the time evolution of unstable states \cite{marekme}. 
In the case of $s$-wave resonances, the dwell time is more useful than 
the phase time concept as it is free of the singularity 
present in the phase time near threshold. 
We shall discuss this point in section II below. 

In the next section we shall first briefly introduce 
the concepts of dwell and phase times and the relations relevant in the 
present work. In section III the formalism used by 
Kapur and Peierls (K-P) is briefly presented and 
its connection with the dwell time of 
Smith and the closely connected definition of phase time and 
Wigner's phase time delay \cite{wigner} is discussed. In section IV, 
the 3-body dwell time 
($\tau^{3-b}$) will be derived using two different approaches. 
The first one uses an extension of the K-P formalism to the 3-body case. 
The second one starts with the standard definition of a dwell time involving 
a 3-body wave function and current density. This derivation leads to a 
3-body dwell time relation given in terms of the component two-body dwell 
times, exactly as obtained within the K-P approach. In section V we 
summarize our findings. 

\section{Dwell and phase time} 
The average 
dwell time for an arbitrary barrier $V(x)$ in one dimension (a framework 
which is also suitable for spherically symmetric problems) confined to an 
interval ($x_1$, $x_2$) is given by the number of particles in the 
region divided by the incident flux $j$: 
\begin{equation}\label{dwellt}
\tau_D(E)\, =\, {\int_{x_1}^{x_2} \, |\, \Psi(x)\,|^2\, dx \over j}\,\, . 
\end{equation} 
Here $\Psi(x)$ is the time independent solution of the Schr\"odinger equation 
in the given region. The dwell and phase time are closely connected and 
for a particle of energy $E\, = \, \hbar^2\, k^2/ 2 \mu$ ($\hbar \,k$ is the 
momentum), incident on the barrier \cite{winfulprl}, 
\begin{equation}\label{dwelphast}
\tau_{\phi} (E)\, =\, 
\tau_D(E)\, - \, \hbar\,{{\rm Im}\,R \over k} \, 
{dk\over dE} \, ,
\end{equation}
where the phase time $\tau_{\phi} (E)$ is given in terms of a weighted sum 
of the energy derivative of the reflection and transmission phases. 
The phase time is essentially the time difference between the arrival and
departure of a wave packet at the barrier. 
$R$ is the reflection coefficient and the second term on the right hand side 
arises due to the interference between the incident and reflected 
waves in front of the barrier. 
This term is important at low energies and
becomes singular as $E \to 0$, thus making the phase time singular too. 
In the case of scattering in three dimensions, the above relation is 
replaced by a very similar one, namely \cite{meprl}, 
\begin{equation}\label{fin}
\tilde{\tau}_D(E) \,=\, \tilde{\tau}_{\phi}(E)\, +\, \hbar \,
\mu\, [t_R / \pi] \,\,dk / dE\,\, ,
\end{equation}
where $t_R$ is the real part of the scattering transition matrix and 
$\mu$ the reduced mass of the two scattering particles. 
$\tilde{\tau}_D(E)$ and $\tilde{\tau}_{\phi}(E)$ are now the dwell and 
phase time ``delays" given by 
$\tilde{\tau}_{\phi}(E)\, =\, \tau_{\phi}(E)\,-\,\tau^0(E) $ and
$\tilde{\tau}_D(E) \,=\,\tau_D(E)\,-\,\tau^0(E)$, with $\tau^0(E)$ 
being the time spent in the same region of space without interaction (or 
in the absence of barrier). $\tilde{\tau}_{\phi}(E)$ is more commonly 
known as Wigner's time delay \cite{wigner} and is given by the energy 
derivative of the scattering phase shift, $\tilde{\tau}_{\phi}(E) \, 
=\, 2 \hbar$ d$\delta$/dE. 

The phase time delay has been found to be very
useful in characterizing resonances in hadron scattering \cite{ourothers}. 
However, due to the singularity mentioned above, the phase time delay  
poses a serious problem in identifying the $s$-wave resonances occurring 
close to threshold. In these particular cases, the dwell time delay 
emerges as the more useful concept \cite{portme, meprl, marekme} since 
it also has a physical significance of being connected to the density 
of states (DOS). A relation between the dwell time and the DOS 
for a system in three dimensions with arbitrary shape 
was derived in \cite{iancon} and \cite{gaspa} discussed 
the same with the example of a symmetric barrier in one dimension. 
The dwell time delay displays the correct threshold behaviour expected 
for the density distribution of an unstable state formed as an 
intermediate state in two body scattering \cite{portme, marekme}. 
We shall now see how the above definitions 
appear in the Kapur-Peierls (K-P) formalism. 

\section{Kapur Peierls formalism revisited} 
In an attempt to obtain a dispersion formula for nuclear reactions, K-P 
considered first the scattering of one particle in a central field of 
force, assuming this field to be fully contained within a sphere of 
radius $r_0$. The partial wave $\phi$ with only $l = 0$ was taken 
into account. $\phi \, =\, r\, \Psi$ satisfies the radial equation \cite{kppl}, 
\begin{equation}\label{kp1}
( \, E \, - \, H\, )\, \phi\, = \, {\hbar^2 \over 2 m} \, {d^2\phi \over 
dr^2 } \, + \, [\, E \,-\, V(r)\, ] \, \phi \, =\, 0 \, \, .
\end{equation}
For $r \ge r_0$, $V(r)$ vanishes and 
\begin{equation}
{d^2\phi \over dr^2} \, +\, k^2\, \phi\, =\, 0 \, \, .
\end{equation}
The solution of this equation is written as, 
$\phi \, = \, (I/k) \, {\rm sin}(kr)\, + \, S\, e^{ikr}$, with $I$ and $S$ 
the amplitudes of the incident and scattered waves. At $r = r_0$, 
\begin{eqnarray}\label{kp3} 
I \, e^{-ikr_0}\, =\, \biggl ( {d\phi \over dr} \biggr )_{r_0} \, - \, 
i \, k\, \phi(r_0) \\ \nonumber
S \, =\, {\rm cos}(k r_0)\, \phi(r_0)\, -\, {1 \over k} \, 
{\rm sin}(k r_0)\, \biggl ( {d\phi \over dr} \biggr )_{r_0}\,\, .
\end{eqnarray}
At this point, K-P impose a boundary condition and obtain a discrete set 
of complex energy eigenvalues. They consider a situation where no incident 
waves are present, which gives rise to the boundary condition,  
\begin{equation}\label{kpbound}
{d\phi \over dr} \, - \, i\,k\,\phi \, =\, 0\,\,\,\,\, ({\rm at}\, \,
r \, =\, r_0) \, \, .
\end{equation}
The boundary condition is obviously not compatible with (\ref{kp1}) 
if $E$ is real, but 
is rather satisfied by the solutions $\phi_n$ such that, 
\begin{equation}\label{kp4}
{\hbar^2 \over 2m} \, {d^2\phi_n \over dr^2} \, +\, [ \, W_n \, -\, V(r)\,] 
\, \phi_n\, = \, 0 \, ,
\end{equation} 
where $V(r)$ is real but $W_n$ are complex eigenvalues. Multiplying 
(\ref{kp4}) by $\phi_n^*$, subtracting the complex conjugate of this 
equation and integrating gives, 
\begin{equation}\label{kp5} 
{\hbar^2 \over 2\, m\, i} \, \biggl [\, \phi_n^*\, {d\phi_n \over dr} 
\, -\, \phi_n\, {d\phi_n^* \over dr} \, \biggr ] \,\biggl|_{r_0} 
\, =\, - 2\, {\rm Im}(W_n)\, \int_0^{r_0}\, \phi_n^*\, \phi_n\, dr\, \, .
\end{equation}
Identifying the left hand side of the above equation with
$\hbar j$, with $j$ being the standard 
quantum mechanical definition of current density at $r_0$ and considering 
$W_n \, =\, E_n\, - \,i \,\Gamma_n/2$, a typical pole of an unstable 
state with width $\Gamma_n$, 
\begin{equation}\label{kp6} 
{\hbar \over \Gamma_n}\, =\, {\int_0^{r_0} \, \phi_n^* \, \phi_n\, dr 
\over j(r_0)} \, \, .
\end{equation}
The right hand side is indeed similar to the definition of dwell time as in 
Eq. (\ref{dwellt}). K-P did not identify the relation with a dwell time 
definition as is now known in literature. Note however that the current 
density appearing in Eq. (\ref{kp6}) is not the incident current density 
but rather the current density at $r_0$. In this sense, Eq. (\ref{kp6}) 
could be compared with a transmission dwell time \cite{wealphagoto} 
rather than an average dwell time as in (\ref{dwellt}). The 
transmission dwell time involves the transmitted current density and 
was shown in \cite{wealphagoto} to be related to the lifetimes of 
unstable nuclei.

The boundary condition (\ref{kpbound}) has interesting 
implications. Indeed, in connection with the work of Smith \cite{smith}, 
Wigner 
remarked \cite{magic} that the time delay 
in \cite{smith} should have been calculated using 
only the outgoing part of the scattered wave. We look at this 
possibility now. 
Using only the outgoing part of the scattered waves 
one would expect the second term on the right 
hand side in Eq. (\ref{dwelphast}) which arises due to the overlap of the 
incident and reflected waves to vanish and $\tau_D(E)\, =\, \tau_{\phi}(E)$. 
To demonstrate the above, we 
repeat the steps in Smith's \cite{smith} derivation of dwell time 
delay with the asymptotic wave function given by 
$_{\infty}\Psi\, =\, (1/\sqrt{v})\, 
[e^{2 i \delta}\, e^{ikx}]$ ($v$ is the velocity) 
instead of the full incident plus scattered wave function,  
$_{\infty}\Psi \, = \, (1/\sqrt{v})\, [e^{-ikx} \, - \, 
e^{2 i \delta}\, e^{ikx}]$. For a wave function $\Psi$ which satisfies 
the Schr\"odinger equation, it is easy to see that,
\begin{equation}
\Psi^* \, \Psi \, =\, - {\hbar^2 \over 2m} \, {\partial \over \partial x} \, 
\biggl (\, \Psi^* {\partial ^2 \Psi \over \partial x \partial E} \, - \, 
{\partial \Psi \over \partial E} {\partial \Psi^* \over \partial x} \, 
\biggr ) \, .
\end{equation}
Since $\Psi^*$ and $\partial \Psi / \partial E$ vanish at $x = 0$, integration 
from $0$ to $r_0$ gives, 
\begin{equation}
\int_0^{r_0} \,\Psi^* \, \Psi \, =\, 
- {\hbar^2 \over 2m} \,\, 
\biggl (\, \Psi^* {\partial ^2 \Psi \over \partial x \partial E} \, - \, 
{\partial \Psi \over \partial E} {\partial \Psi^* \over \partial x} \, 
\biggr )_{r_0}\, .
\end{equation}
At large $r_0$, we replace $\Psi$ on the right hand side of the above equation 
by $_{\infty}\Psi$ and obtain, 
\begin{equation}
\int_0^{r_0}\, \Psi^* \, \Psi\, dx \, -\, {r_0\over v} \, =\, 2 \hbar 
\, {d\delta \over dE}\,\,. 
\end{equation}
With $r_0/v \,=\, \tau^0$, the time spent without interaction, the 
left hand side of the above equation can be identified with a dwell time 
delay and since $2 \hbar \,d\delta /dE\, =\, \tilde{\tau}_{\phi}(E)$, 
we get $\tilde{\tau}_D(E) \, =\, \tilde{\tau}_{\phi}(E)$. 
We note in passing that the wave number $k$ appearing in the above equations 
is real and is the same for all complex eigenvalues $W_n$. This is a result of 
the boundary condition (\ref{kpbound}). 
In contrast to this approach, Gamow \cite{gamo} introduced 
standing waves in front of the barrier with the result that the 
asymptotic outgoing wave is a plane wave with a complex wave number $k$. 
Other approaches which deal with solutions of the Schr\"odinger 
equation for complex energies can be found in \cite{kukulin}. 

{\it New definition of dwell time}

Starting with the scattering amplitude as given in Eq.(\ref{kp3}), 
namely, 
$S \, =\, {\rm cos}(k r_0)\, \phi(r_0)\, -\, {1 \over k} \, 
{\rm sin}(k r_0)\, ( d\phi /dr)_{r_0}$ and noting the standard definition 
of $S\, =\, e^{2i\delta}$ (where $\delta$ is the scattering phase shift), 
we can write $e^{2 i \delta} \, =\, \phi(r_0)\, e^{-i k r_0}$. Taking the 
energy derivative of this equation, it is easy to see that, 
\begin{equation}\label{newd1} 
2 \, \hbar \, {d\delta \over dE} \, +\, {r_0 \over v} \, =\, 
-\, i\, \hbar \, {1 \over \phi(r_0)}\,{d\phi(r_0)\over dE}\, \,, 
\end{equation}
with $v\, =\, \hbar k /m$. $r_0/v \, =\, \tau^0$ is the time spent without 
interaction in the region of radius $r_0$. 
$2 \, \hbar \, d\delta/dE$ is the phase time delay 
($2 \, \hbar \, d\delta/dE \, =\, \tilde{\tau}_{\phi}(E)\, =\, 
\tau_{\phi}(E)\,-\,\tau^0(E)$) and hence 
the left hand side of (\ref{newd1}) is simply $\tau_{\phi}(E)$. 
If the boundary condition (\ref{kpbound}) is imposed, we have 
already seen that $\tau_{\phi}(E)\,=\, \tau_D(E)$ and from the equations 
above, 
\begin{equation}
\tau_D(E) \, =\, -\, i\, \hbar \, {d\over dE}[\,{\rm ln}\,\, \phi(r_0)\,]
\, \, ,
\end{equation}
which is a new definition of dwell time obtained within the K-P formalism. 
\section{Three body dwell time}
Since the dwell and phase time concepts have been successfully 
used \cite{wealphagoto, meprl, ourothers} to study resonances occurring in
two body elastic scattering, it appears timely to extend these ideas
for the study of unstable systems which can be viewed upon as three body
systems. Such unstable states occur in different branches of physics. 
For example, in a recent study \cite{garrido} of the $s$-wave resonances 
$^9$Be and $^9$B, it was shown that these unstable nuclei can be looked 
upon as genuine three body resonances with the $^9$Be for example being 
composed of the substructure $\alpha$\,+\,$\alpha$ + n. Other examples
could include hadronic systems of two mesons and a baryon, 
two neutron halo nuclei or even hypernuclei such as 
$^6$He$_{\Lambda}$ (with substructure $^4$He \, +\, 
$\Lambda$ \, +\, n). Since this work is a first attempt 
to derive an expression 
for the dwell time of such three body systems, we restrict ourselves to
the case of $s$-waves, i.e. we consider only the partial wave with $l = 0$. 
This allows us to develop the formalism in analogy to the dwell time 
formalism in the one dimensional case. We shall further assume that the 
wave function can be expressed in a separable form (which is often also the 
case in studies of three body systems mentioned above). 
\subsection{Lifetime of a three body resonance} 
To describe the three particle system, we start by writing the Hamiltonian 
as \cite{joach}, 
\begin{equation}\label{3bodyschrod} 
H \, =\, {{\bf p}^2 \over 2 \mu_1} \, +\, {{\bf q}^2 \over 2 \mu_2} \, +\, 
V^1(\mbox{\boldmath$\rho$})\, +\, 
V^2\biggl (\,{\bf r} \, +\, {m_2 \over m_2 + m_3} \, 
\mbox{\boldmath$\rho$}\,\biggr ) 
\, +\, V^3\biggl (\,{\bf r}\, -\, {m_3 \over m_2 + m_3} \, 
\mbox{\boldmath$\rho$}
\, \biggr )\, \, , 
\end{equation}
where ${\bf q}$ is relative momentum of particles $2$ and $3$ and ${\bf p}$ 
that of 
particle $1$ and the compound system made up of ($2,3$). These are conjugate 
momenta to the position vectors, 
\begin{equation} 
{\bf r} \, =\, {\bf r}_1 \, - \, {m_2\, {\bf r}_2 \, +\, m_3\, {\bf r}_3 
\over m_2 \, +\, m_3} \, ,\,\, \, \, \, \, \mbox{\boldmath$\rho$} 
\, =\, {\bf r}_2\, - 
\, {\bf r}_3 \,\,.
\end{equation} 
In principle, there can be three such sets of coordinates 
({\bf r}, $\mbox{\boldmath$\rho$}$) depending on the choice of the 
subsystems. We now operate $H$ in (\ref{3bodyschrod}) on the wave function 
$\Psi({\bf r}, $\mbox{\boldmath$\rho$}$)$ assuming a separable form for 
this wave function. 
As mentioned above, we shall restrict to spherical symmetry ($s$-waves) 
and hence retain only the radial part of both the {\bf r}
and  $\mbox{\boldmath$\rho$}$ 
coordinates in the above equation. 
Thus, writing $\Psi({\bf r}, \mbox{\boldmath$\rho$}) \, =\, F({\bf r})\, 
G(\mbox{\boldmath$\rho$})$, with $F(r) \,=\, \chi(r)/r$, 
$G(\rho)\,=\, \Phi(\rho)/\rho$ and using 
$${\bf p}^2 \,=\, -\, \hbar^2\, {1 \over r^2}\, {d \over dr}\biggl(r^2\, 
{d\over dr}\,\, \biggr )\,\, ,$$
$${\bf q}^2 \,=\, -\, \hbar^2\, {1 \over \rho^2}\, {d \over d\rho}\biggl(
\rho^2\, {d\over d\rho}\,\, \biggr )\,\, ,$$ one obtains  
in analogy to (\ref{kp1}):
\begin{equation}
E \, \chi(r)\,\Phi(\rho) \, +\, \Phi(\rho)\, {\hbar^2 \over 2 \mu_1}\, 
{d^2\chi(r)\over d r^2} \, +\, 
 \chi(r)\, {\hbar^2 \over 2 \mu_2}\, 
{d^2\Phi(\rho)\over d\rho^2} \, -\,\tilde{V}\,\chi(r)\, 
\Phi(\rho)\, =\, 0\, \, , 
\end{equation}
where $\tilde{V}$ is the sum of the three potentials and $E$ the energy 
eigenvalue of the three body system. 
From this point we continue as in the K-P formalism where the potentials 
are real and the complex energy, $E\, =\, E_R \, - \, i\, \Gamma_R/2$.   
We now multiply 
the above equation by $\chi^*(r)\, \Phi^*(\rho)$, take the complex conjugate 
of the resulting equation and subtract it from the original one and then 
integrate the resulting equation over the radial coordinates $r$ and $\rho$ 
to obtain, 
\begin{equation}
- \,i\, \, N_{\chi}\, N_{\Phi}\, \Gamma_R\, +\, N_{\Phi}\,
{\hbar^2 \over 2 \mu_1}\, 
\biggl [\, \chi^*\, {d\chi \over dr} 
\, -\, \chi\, {d\chi^* \over dr} \, \biggr ] \,\biggl|_{r_{\chi}} \, 
+\,  N_{\chi}\, 
{\hbar^2 \over 2 \mu_2}\,
\biggl [\, \Phi^*\, {d\Phi \over d\rho} 
\, -\, \Phi\, {d\Phi^* \over d\rho} \, \biggr ] \,\biggl|_{\rho_{\Phi}} \, 
= \, 0\, \, ,
\end{equation}
where $N_{\chi}\, =\, \int_0^{r_{\chi}} \, |\chi|^2\, dr$ and 
$N_{\Phi}\, =\, \int_0^{\rho_{\Phi}} \, |\Phi|^2\, d\rho$. 
Dividing throughout by $N_{\chi}\, N_{\Phi}\,/ (\hbar i)$ and identifying 
\begin{eqnarray}  
j_{\chi} \, =\, {\hbar \over 2 \,\mu_1\, i}\, 
\biggl [\, \chi^*\, {d\chi \over dr} 
\, -\, \chi\, {d\chi^* \over dr} \, \biggr ] \,\biggl|_{r_{\chi}} \\ 
\nonumber
j_{\Phi} \, =\, {\hbar \over 2 \,\mu_2\, i}\,
\biggl [\, \Phi^*\, {d\Phi \over d\rho} 
\, -\, \Phi\, {d\Phi^* \over d\rho} \, \biggr ] \,\biggl|_{\rho_{\Phi}}\,
\end{eqnarray}
as the quantum mechanical current densities, we get, 
\begin{equation}
{\Gamma_R \over \hbar} \, =\, {j_{\chi} \over 
 \int_0^{r_{\chi}} \, |\chi|^2\, dr} \, +\, {j_{\Phi} \over 
\int_0^{\rho_{\Phi}} \, |\Phi|^2\, d\rho}\, \, .
\end{equation}
Applying the definition (\ref{dwellt}) to the right hand side, 
the lifetime of the three body system $\tau^R \, =\, \Gamma_R/\hbar$ 
is thus given in terms of the two-body dwell times as, 
\begin{equation}
{1 \over \tau^R} \, =\, {1 \over \tau_D^{\chi}}\, +\, 
{1 \over \tau_D^{\Phi}}\, \, ,
\end{equation}
where $\tau_D^{\Phi}$ is the time spent by particles $2$ and $3$ within 
a spherical region of radius $\rho_{\Phi}$ and $\tau_D^{\chi}$ is the time 
spent by particle $1$ and the composite system $(2,3)$ within a sphere 
of radius $r_{\chi}$. Below we shall see that the $\tau^R$ derived above 
is indeed the definition of a three body dwell time. At this point it is 
nice to note that such an inverse addition of dwell times was also found 
in \cite{wealphagoto} in connection with the reflection, transmission and 
average dwell times for a particle tunneling a barrier. 

\subsection{Three body current density and dwell time} 
Following the standard definition of the dwell time as in (\ref{dwellt}) 
one can define the three body dwell time in terms of a quantum mechanical 
current density for a three body system as: 
\begin{equation}\label{dwelt3} 
\tau^{3-b} \, =\, {\int_0^{r_{\chi}}\, \int_0^{\rho_{\Phi}}\, \, 
|\, \Psi\, |^2\, 
dr\, d\rho \over j_{3-b}}\, \, ,
\end{equation}
where $j_{3-b}$ is the three body current density and 
$\Psi$ the three body wave function which we write in terms of separable 
wave functions below. Though one would 
guess $j_{3-b}$ to be a sum over the already defined $j_{\chi}$ and 
$j_{\Phi}$, if one tries to derive a continuity equation for the 
three body system, one finds that this is indeed not the case. 
In general, the 
non-relativistic definition of a many body current density is 
not trivial and has been dealt with using different 
approaches \cite{feynman}. Here, we start with an approach 
(see Appendix) used for an 
$N$-body system with wave function $\Psi({\bf r}_1,{\bf r}_2,{\bf r}_3 ...
{\bf r}_N)$ and 
define the current densities in terms of the coordinates 
({\bf r}, $\mbox{\boldmath$\rho$}$) instead of (${\bf r}_1,{\bf r}_2,{
\bf r}_3$). 
For the separable 
wave function $\Psi(r,\rho,t)\,=\, \chi(r)\, \Phi(\rho)\, f(t)$ 
(with $l = 0$ only) which satisfies the 
Schr\"odinger equation $H \Psi\,=\, i \hbar \partial \Psi/\partial t$, 
we can write, 
\begin{eqnarray}
{\partial |\Psi|^2 \over \partial t} & =& {1 \over i \hbar} \, 
(\Psi^* \, H\, \Psi\, -\, \Psi\, H\, \Psi^*)\\ \nonumber 
&=& - {\hbar \over 2 \mu_1 i} \, |\Phi|^2  \, |f|^2\, 
\, {d\over dr}\, \, \biggl (\chi^*\,{d\chi\over dr}\, -\, 
\chi\, {d\chi^*\over dr} \,\biggr)
\, -\, 
{\hbar \over 2 \mu_2 i} \, |\chi|^2  \, |f|^2\, 
{d\over d\rho} \, \biggl(\Phi^*\,{d\Phi\over d\rho}\, - \, 
\Phi\, {d\Phi^* \over d\rho}\,\biggr) \\ \nonumber
&=& - {\hbar \over \mu_1}\, |\Phi|^2  \, |f|^2\, \nabla_r \,{\rm Im} 
\, \biggl (\chi^*\,{d\chi\over dr} \, \biggr )\, -\, 
{\hbar \over \mu_2}\, |\chi|^2  \, |f|^2\, \nabla_{\rho} \,{\rm Im} 
\, \biggl (\Phi^*\,{d\Phi\over d\rho} \, \biggr )\, \, ,
\end{eqnarray} 
implying, 
\begin{equation} 
{\partial |\Psi|^2 \over \partial t} \, +\, \nabla_r \, J_r\, +\, 
\nabla_{\rho}\,J_{\rho} \,=\, 0\,\, , 
\end{equation}
where we denote $\partial/\partial r \, =\,\nabla_r$ and 
$\partial/\partial \rho \, =\,\nabla_{\rho}$. 
With $J_r\, =\, |\Phi|^2  \, |f|^2\,j_{\chi}$ and 
$J_{\rho}\, =\, |\chi|^2  \, |f|^2\,j_{\Phi}$,  
the above equation has the form $\partial |\Psi|^2/\partial t \, +\, 
\sum_{i =1,2}\, \nabla_i \,J_i\, =\,0$ (with $(1, 2)$ 
corresponding to ($r,\rho$)) 
and is not a continuity equation of the 
standard form. Hence, we rather define new current densities, $j_r$ and 
$j_{\rho}$ (in analogy to those explained in the appendix), such that, 
\begin{eqnarray}\label{currents}
j_r \, =\, {\hbar \over \mu_1}\,|f|^2 \, N_{\Phi}\, {\rm Im} \, 
\biggl ( \chi^* \, {\partial \chi \over \partial r} \, \biggr ) \,=\, 
|f|^2\, N_{\Phi}\, j_{\chi}\\ \nonumber 
j_{\rho} \, =\, {\hbar \over \mu_2}\,|f|^2 \, N_{\chi}\, {\rm Im} \, 
\biggl ( \Phi^* \, {\partial \Phi \over \partial \rho} \, \biggr ) \,=\, 
|f|^2\, N_{\chi}\, j_{\Phi}\, \, .
\end{eqnarray} 
It can be easily checked that the above current densities 
satisfy the individual continuity equations, $\partial n_r/ \partial t\,=\, 
-\, \nabla_r\, j_r$ and $\partial n_{\rho}/ \partial t\,=\, 
-\, \nabla_{\rho}\, j_{\rho}$, with 
$n_r\, =\, N_{\Phi} \,|\chi|^2\, |f|^2$ and 
$n_{\rho}\, =\, N_{\chi} \,|\Phi|^2\, |f|^2$. 

Replacing the current density $j_{3-b}\, =\, j_r \, +\,j_{\rho}$ in 
(\ref{dwelt3}) along with the definitions (\ref{currents}) 
and using the separable form of the wave function 
as before, it is easy to verify that, 
\begin{eqnarray}\label{dwelt32} 
{1 \over \tau^{3-b}} & =& {j_{\chi} \over N_{\chi}}\, +\, {j_{\Phi} 
\over N_{\Phi}} \\ \nonumber
& = & {1 \over \tau_D^{\chi}}\, +\, {1 \over \tau_D^{\Phi}} \, \, .
\end{eqnarray} 
We have thus shown that the lifetime of the resonance $\tau^R$ derived 
previously is the same as $\tau^{3-b}$ which can be expressed in terms
of the two body dwell times, $\tau_D^{\chi}$ and $\tau_D^{\Phi}$. 

\section{Summary}
There exist several concepts and definitions of quantum tunneling times
in literature. However, among these, the dwell time concept seems to be 
one of the most important concept considering the variety of applications 
it finds in different branches of physics as mentioned in the introduction. 
In the present work we make a first attempt to find a relation for the 
dwell time of a three body system. 

The findings of the present work can be summarized point wise as:
\begin{enumerate}
\item {The dwell time relation derived by Kapur and Peierls (K-P) is 
rediscovered and yet another new expression for the dwell time 
is obtained within 
their formalism. Within the approach used by Kapur and Peierls, the dwell 
time is shown to be equal to the phase time and hence also
free of any singularities near threshold.}
\item {Using a similar formalism as that of K-P for a three body resonance, 
the lifetime of such a resonance is shown to be related to the 
two body dwell times of the substructures of the three body system. }
\item {Starting from the standard definition of dwell time taken along 
with a three body current density it is shown that the three body dwell 
time is exactly equal to the lifetime found in (2.) and is thus related 
to the two body dwell times of the substructures. }
\end{enumerate} 
Though the approach of the present work relies on 
simplistic assumptions of spherically symmetric potentials and separable 
wave functions (which do find applications in certain physical examples), 
the results obtained are interesting and motivating enough 
to continue further investigations of the three body dwell time concept. 

\renewcommand{\theequation}{A-\arabic{equation}}
\setcounter{equation}{0}
\section*{APPENDIX: Probability and current densities in many body 
non-relativistic quantum mechanics}

Given the wave function $\Psi({\bf r}, t)$ for a quantum system which 
satisfies the one-body Schr\"odinger equation $H \, \Psi\,=\, i \, \hbar \, 
\partial\Psi/\partial t$, with $H \,=\, -(\hbar^2/2m)\, \nabla^2\, +\, 
V({\bf r})$, one can derive the continuity equation 
\begin{equation}\label{continuity}
{\partial |\Psi|^2 \over \partial t} \, =\, -\, \mbox{\boldmath$\nabla$} 
\cdot {\bf J}\, \, ,
\end{equation}
where {\bf J} is the current density given by ${\bf J} \, =\,(\hbar/m) \, 
{\rm Im}\, (\Psi^*\, \mbox{\boldmath$\nabla$} \, \Psi)$. 
Consider now a many body system consisting of $N$ 
particles and described by the wave function $\Psi({\bf r}_1, 
{\bf r}_2, {\bf r}_3, {\bf r}_4, ......., {\bf r}_N,t)$ which satisfies  
the Schr\"odinger equation $H \, \Psi\,=\, i \, \hbar \, 
\partial\Psi/\partial t$, however, with 
\begin{equation}\label{manybodyh}
H\, =\, - {\hbar^2 \over 2} \, \sum_{i\,=\,1}^N\, \, {1 \over m_i} \, \, 
\mbox{\boldmath$\nabla$}_i^2 \,\, +\, V ({\bf r}_1, 
{\bf r}_2, {\bf r}_3, {\bf r}_4, ......., {\bf r}_N )\, \, .
\end{equation}  
Starting in the standard way, with, 
\begin{equation}
{\partial |\Psi|^2 \over \partial t}\, =\, {1 \over i\hbar} \, 
(\, \Psi^*\, H\, \Psi\, -\, \Psi\, H\, \Psi^*\, )\, \, ,
\end{equation}
and using the Hamiltonian in (\ref{manybodyh}), one obtains,
\begin{equation}\label{continuity3}
{\partial |\Psi|^2 \over \partial t}\, =\,-\,\hbar\, 
\sum_{i\,=\,1}^N\, \, {1 \over m_i}\,\mbox{\boldmath$\nabla$}_i \cdot 
{\rm Im} (\Psi^*\, \mbox{\boldmath$\nabla$}_i\, \Psi ) \, =\, 
-\, \sum_{i\,=\,1}^N\,\mbox{\boldmath$\nabla$}_i \cdot {\bf J}_i\, \, , 
\end{equation}
where we have defined the current density for particle $i$ as 
${\bf J}_i({\bf r}_i,t)\,=\, (\hbar / m_i) \, 
{\rm Im} (\Psi^*\, \mbox{\boldmath$\nabla$}_i\, \Psi )$. Eq. 
(\ref{continuity3}) is obviously not of the same form as (\ref{continuity}) 
with ${\bf J}\,=\,\sum_i\,{\bf J}_i$. Hence, one defines \cite{feynman} the 
number and current density rather for particle $1$ as, 
\begin{eqnarray}
n_1 ({\bf r},t)\, =\, \int\, d{\bf r}_2\,d{\bf r}_3 \, ...\, d{\bf r}_N\, 
|\Psi|^2 \, \, ,\\ \nonumber
{\bf j}_1\, =\, {\hbar \over m_1} \, \, \int\, d{\bf r}_2\,d{\bf r}_3 \, 
...\, d{\bf r}_N\,{\rm Im} (\Psi^*\, \mbox{\boldmath$\nabla$}_1\, \Psi) \, \, 
\end{eqnarray} 
and so on for other particles too. It can be easily checked \cite{feynman} 
that, particle $1$ 
satisfies the continuity equation 
$\partial n_1/ \partial t\, =\, 
\mbox{\boldmath$\nabla$}_1 \cdot {\bf j}_1$ and so does every particle in 
the system of $N$ particles.

In analogy to the above procedure, we define the current densities for 
the three body system. However, instead of the position coordinates 
$({\bf r}_1, {\bf r}_2, {\bf r}_3)$ of the three particles, we use the 
coordinates ${\bf r}$ and $\mbox{\boldmath$\rho$}$ described in the text. 
Thus we define the two current densities: 
\begin{eqnarray}
j_r\,=\, {\hbar \over \mu_1}\, \, \int\, d\mbox{\boldmath$\rho$}\, 
{\rm Im} [\,\Psi^*({\bf r},\mbox{\boldmath$\rho$}, t) 
\, \mbox{\boldmath$\nabla$}_r\, \Psi({\bf r},\mbox{\boldmath$\rho$}, t)\,]
\\ \nonumber
j_{\rho}\,=\, {\hbar \over \mu_2}\, \, \int\, d{\bf r}\, 
{\rm Im} [\,\Psi^*({\bf r},\mbox{\boldmath$\rho$}, t) 
\, \mbox{\boldmath$\nabla$}_{\rho}\, \Psi({\bf r},\mbox{\boldmath$\rho$}, t)\,]
\, \, .
\end{eqnarray}
which reduce to a simpler form in case of a separable wave function as 
used in the text.

\end{document}